\documentclass[pra,twocolumn,preprintnumbers,amsmath,amssymb,superscriptaddress]{revtex4}

\usepackage[pdftex]{graphicx, color}
\usepackage{amsmath}
\usepackage{amssymb}
\usepackage{amsthm}
\usepackage{amsfonts}						%
\usepackage{graphicx}
\usepackage[english]{babel}             	
\usepackage{epstopdf}
\usepackage{graphicx}                   	

\begin{document}

\title{Signature of a three-dimensional photonic band gap observed on silicon inverse woodpile photonic crystals}

\author{Simon R. Huisman} \email{s.r.huisman@utwente.nl, www.photonicbandgaps.com}
\affiliation{Complex Photonic Systems (COPS), MESA+ Institute for
Nanotechnology, University of Twente, PO Box 217, 7500 AE Enschede, The Netherlands}
\author{Rajesh V. Nair}
\affiliation{Complex Photonic Systems (COPS), MESA+ Institute for
Nanotechnology, University of Twente, PO Box 217, 7500 AE Enschede, The Netherlands}
\author{L\'eon A. Woldering}
\affiliation{Complex Photonic Systems (COPS), MESA+ Institute for
Nanotechnology, University of Twente, PO Box 217, 7500 AE Enschede, The Netherlands}
\author{Merel D. Leistikow}
\affiliation{Complex Photonic Systems (COPS), MESA+ Institute for
Nanotechnology, University of Twente, PO Box 217, 7500 AE Enschede, The Netherlands}
\affiliation{Center for Nanophotonics, FOM Institute for Atomic and
Molecular Physics (AMOLF), Science Park 113, 1098 XG Amsterdam,
The Netherlands}%
\author{Allard P. Mosk}
\affiliation{Complex Photonic Systems (COPS), MESA Institute for
Nanotechnology, University of Twente, PO Box 217, 7500 AE Enschede, The Netherlands}
\author{Willem L. Vos}
\affiliation{Complex Photonic Systems (COPS), MESA+ Institute for
Nanotechnology, University of Twente, PO Box 217, 7500 AE Enschede, The Netherlands}

\date{\today}

\begin{abstract}
We have studied the reflectivity of CMOS-compatible three-dimensional silicon inverse woodpile photonic crystals at near-infrared frequencies. Polarization-resolved reflectivity spectra were obtained from two orthogonal crystal surfaces using an objective with a high numerical aperture. The spectra reveal broad peaks with maximum reflectivity of $67\%$ that are independent of the spatial position on the crystals. The spectrally overlapping reflectivity peaks for all directions and polarizations form the signature of a broad photonic band gap with a relative bandwidth up to $16\%$. This signature is supported with stopgaps in plane wave bandstructure calculations and with the frequency region of the expected band gap.

\end{abstract}
\maketitle

\section{Introduction}

Currently, many efforts are devoted to create an intricate class of three-dimensional meta-materials known as photonic crystals that radically control propagation and emission of light \cite{Yablonovitch1987, John1987, Joannopoulos1997, Fleming2002, Koenderink2003, Lopez2003, Lodahl2004, Ogawa2004, Han2007, Takahashi2009, Ergin2010, Tandaechanurat2010}. Photonic crystals are ordered composite materials with a spatially varying dielectric constant that has a periodicity of the order of the wavelength of light \cite{Joannopoulos2008}. As a result of the long-range periodic order, the photon dispersion relations are organized in bands, analogous to electron bands in solids \cite{AshcroftMermin}. Frequency ranges called stopgaps emerge in which light is forbidden to propagate in particular directions due to Bragg interference \cite{Vos1996}. In specific three-dimensional crystals, a common stopgap is formed for all polarizations and directions, called the photonic band gap. Light cannot propagate inside the photonic band gap, allowing for ultimate control of light emission. Emission rates and directions can be manipulated \cite{Koenderink2003, Lodahl2004, Ogawa2004, Tandaechanurat2010}, which could lead to efficient micro-scale light sources \cite{Wierer2009, Tandaechanurat2010} and solar cells \cite{Zhou2008}. Additional interest is aroused by the possibility of Anderson localization of light by point defects added to photonic band gap crystals \cite{John1987}.

It is an outstanding challenge to experimentally demonstrate a photonic band gap; inside the photonic band gap the density of optical states equals zero. The density of states can be investigated with light emitters placed inside the crystal, see \textit{e.g.} Ref.~\cite{Koenderink2003, Lodahl2004, Ogawa2004}. These experiments are difficult to perform and to interpret, and are not always possible due to the limited availability of appropriate light sources and detection methods. However, one can obtain an indication of the band gap by observing a stopband in a directional experiment, such as a peak in reflectivity or a trough in transmission \cite{Fleming1999, Blanco2000, Noda2000, Vlasov2000, Subramania2004, Schilling2005}. The frequency width of an experimentally observed stopband often corresponds to a stopgap in the dispersion relation, where light is forbidden to propagate. Interestingly however, a peak in reflectivity or a trough in transmission also occurs when an incident wave cannot couple to a field mode in the crystal \cite{Robertson1992, Sakoda1995, Joannopoulos2008}. Therefore, one typically compares observed stopbands with theory such as calculated bandstructures to indicate the presence of a band gap. Unfortunately, most theory is typically valid for ideal structures, for instance of infinite size. More compelling evidence of a photonic band gap could be obtained if one demonstrates that stopbands have a common overlap range independent of the measurement direction, since light is forbidden to propagate in a photonic band gap. Angle-resolved \cite{Vos2000, Takahashi2009} and angle-averaged spectra \cite{Schilling2005} have been collected, however, due to experimental considerations it appears to be extremely difficult to measure over $4 \pi$ sr solid angle. Stopbands are typically investigated on only one surface of the photonic crystal, and only few studies have investigated multiple surfaces, see \textit{e.g.} Ref.~\cite{Palacios2002}. In addition, one needs to rule out spurious boundary effects by confirming that stopbands reproduce at different locations on the crystal, requiring position-dependent experiments. Furthermore, it can occur that field modes in the crystal can only couple to a specific polarization \cite{Ho1990}. Polarization-resolved experiments are required to demonstrate that stopbands are present for all polarizations \cite{Staude2011}. Therefore, a strong experimental signature for a photonic band gap is obtained if one can demonstrate that stopbands are position-independent and overlap for different directions and orthogonal polarizations. To the best of our knowledge, such a detailed analysis of stopbands for three-dimensional photonic band gap crystals has not yet been reported.

In this paper we study silicon three-dimensional inverse woodpile photonic crystals \cite{Ho1993, Hillebrand2004}. The inverse woodpile photonic crystal is a very interesting nanophotonic structure on account of its broad theoretical photonic band gap with more than $25 \%$ relative gap width. Schilling \textit{et al.} \cite{Schilling2005} were the first to investigate stopbands in inverse woodpile crystals. They measured unpolarized reflectivity along one crystal direction using an objective with numerical aperture NA = 0.57. An indication for the photonic band gap was found by a stopband that agreed with the calculated band structure. Other groups have fabricated inverse woodpile photonic crystals of different materials using several methods, and also performed reflectivity measurements along one direction with unpolarized light \cite{Santamaria2007, Hermatschweiler2007, Jia2007}, resulting in similar indications of the band gap. Here, we present an extensive set of polarization-resolved reflectivity spectra of silicon inverse woodpile photonic crystals. We have collected spectra from two orthogonal crystal surfaces using an objective with a high numerical aperture NA = 0.65. The reflectivity spectra were obtained on different locations on the photonic crystal surfaces to confirm the reproducibility, to determine the optical size of the crystals, and to investigate boundary effects. We demonstrate position-independent overlapping stopbands for orthogonal polarizations and crystal directions, which is a signature of a three-dimensional photonic band gap. This signature agrees with calculated stopgaps in plane wave bandstructure calculations and with the frequency region of the expected band gap.

\begin{figure}[tbp]
  \includegraphics[width=8.6 cm]{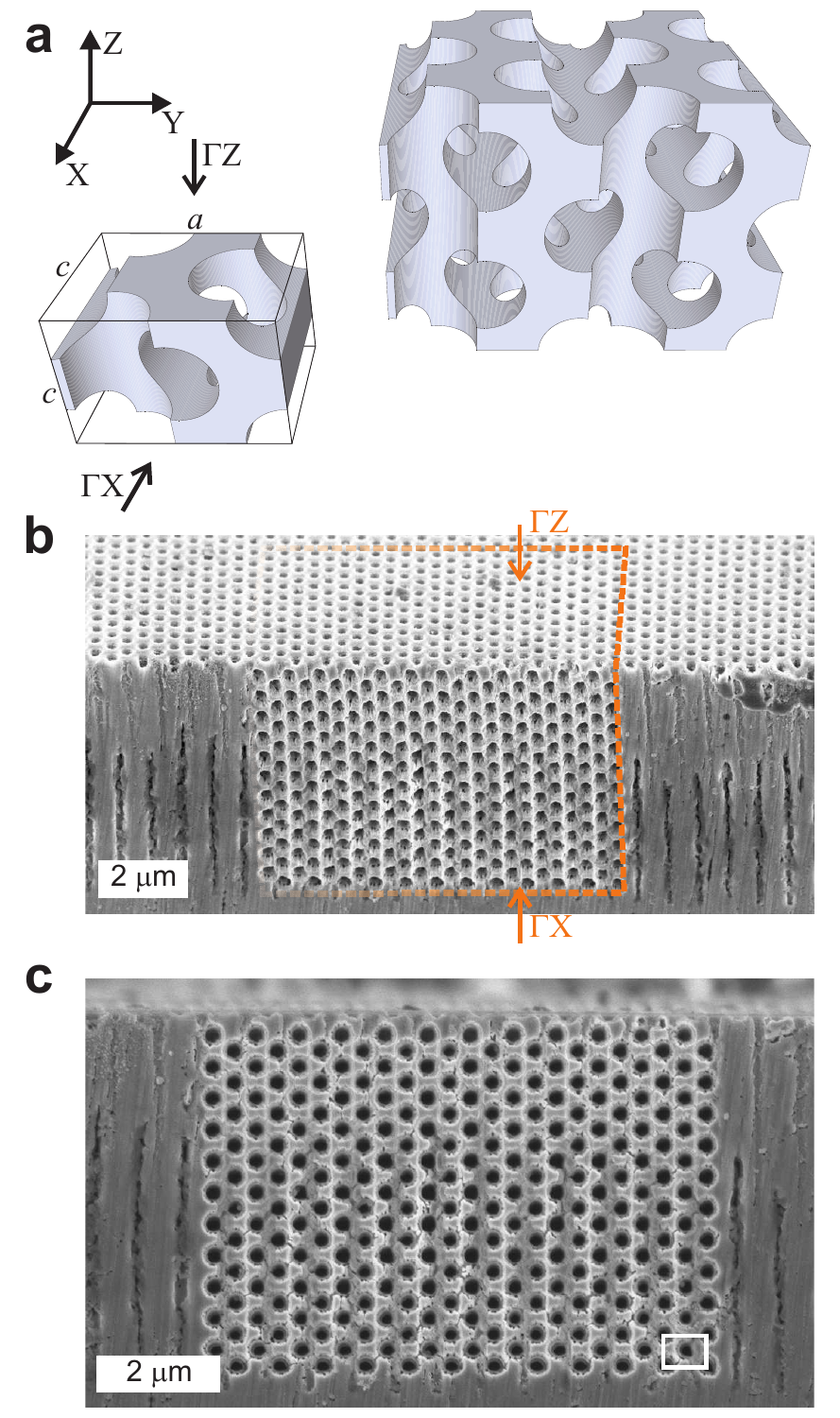}
\caption{\textit{(color online)} Structure of inverse woodpile photonic crystals. $(a)$, Schematic representation of the orthorhombic unit cell of a cubic inverse woodpile photonic crystal (left) and a crystal consisting of eight unit cells (right). The structure can be described by two identical sets of pores running in two orthogonal sides of a box. $(b)$, Scanning electron microscope image of a cubic inverse woodpile crystal with $a = 693 \pm 10$ nm, $c = 488 \pm 11$ nm ($\frac{a}{c} = \sqrt{2}$) and $r=145 \pm 9$ nm, surrounded by a macroporous two-dimensional crystal. The dashed orange line indicates a part of the boundary of the inverse woodpile photonic crystal. $(c)$, Scanning electron microscope image viewed from the $\Gamma X$-direction. The rectangle represents one face of the orthorhombic unit cell, compare with $(a)$.}
\label{fig1}
\end{figure}

\section{Inverse woodpile photonic band gap crystals}
\label{sec2}

Fig. \ref{fig1}$(a)$ illustrates the orthorhombic unit cell of an inverse woodpile photonic crystal (left), together with a crystal consisting of eight unit cells (right). The structure can be described by two identical sets of pores with radius $r$ running in two orthogonal sides of a box, where each set represents a centred-rectangular lattice with sides $a$ and $c$. If the ratio $\frac{a}{c}$ equals $\sqrt{2}$, the crystal symmetry is cubic and the crystal structure is diamond-like, see Ref. \cite{Ho1993, Hillebrand2004, Woldering2009}. This type of inverse woodpile photonic crystals with $\frac{a}{c}=\sqrt{2}$ has a maximum band gap width of $25.4\%$ for $\frac{r}{a}=0.245$. The sets of pores are oriented perpendicular to each other and the centrers of one set of pores are aligned exactly between columns of pores of the other set, resulting in the structure of Fig. \ref{fig1}$(a)$. In fig. \ref{fig1}$(a)$ a coordinate system is introduced that is used in this paper.

We have fabricated multiple inverse woodpile photonic crystals with different pore radii in monocrystalline silicon with a CMOS compatible method, as is described in detail in Ref. \cite{vandenBroek2011, Tjerkstra2011}. Fig. \ref{fig1}$(b)$ shows a scanning electron microscope (SEM) image of one of these crystals from the same perspective as in Fig. \ref{fig1}$(a)$. This crystal consists of more than 10$^3$ unit cells and is located on top of bulk silicon. The crystal is surrounded by a macroporous two-dimensional photonic crystal that forms the first set of pores used to fabricate inverse woodpile photonic crystals. A second set of pores is oriented perpendicular to this two-dimensional crystal to form the inverse woodpile photonic crystal. A part of the boundary of the inverse woodpile photonic crystal is marked with the orange dashed line. The inverse woodpile photonic crystal has lattice parameters $a=693 \pm 10$ nm, $c=488 \pm 11$ nm, and a pore radius of $r=145 \pm 9$ nm in both directions \cite{CallibrationErrors}. Typically eight different inverse woodpile photonic crystals are made on one two-dimensional photonic crystal that each extend over approximately $7$ $\mu$m $\times$ $5$ $\mu$m $\times$ $5$ $\mu$m in size.
Fig. \ref{fig1}$(c)$ shows a SEM image of a crystal viewed parallel with the second set of pores or parallel with the $\Gamma X$-direction. In Ref.~\cite{vandenBroek2011} it has been determined that the second set of pores are precisely centred to within $\Delta y=17 \pm 12$ nm between columns of pores of the first etch direction, which is extremely close to the ideal structure of Fig. \ref{fig1}$(a)$. One unit cell is marked in the right bottom corner of Fig. \ref{fig1}$(c)$. Here, we describe an extensive set of reflectivity measurements collected from this inverse woodpile photonic crystal centred along the $\Gamma X$- and $\Gamma Z$-direction, which is representative for more than five other crystals that we have studied.

\begin{figure}[tbp]
  \includegraphics[width=8.6 cm]{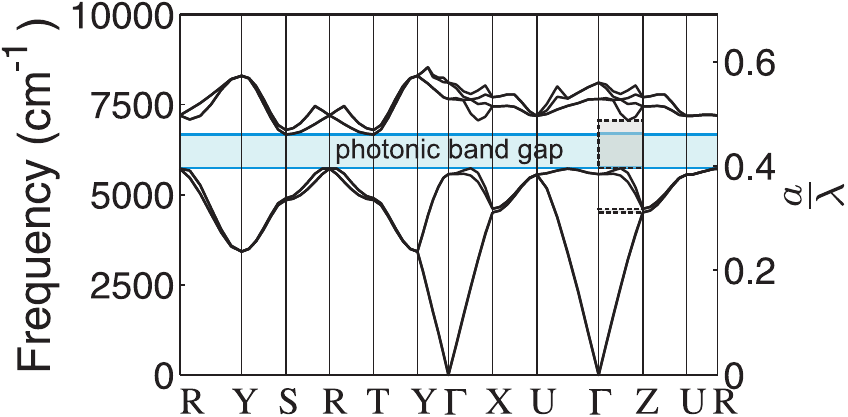}
\caption{\textit{(color online)} Bandstructure of an inverse woodpile photonic crystal. The dispersion relation for the eight lowest frequency Bloch modes are calculated for a structure with $a = 690$ nm, $c = 488$ nm and pore radius $r = 131$ nm ($\frac{r}{a}=0.190$). The left axis represents the absolute frequency in cm$^{-1}$, the right axis shows the reduced frequency. The blue bar marks a photonic band gap with a relative gap width of $\frac{\Delta \omega}{\omega_c}=15.1 \%$. The grey bars mark stopgaps in the $\Gamma Z$-direction.}
\label{fig2}
\end{figure}

Fig. \ref{fig2} shows the bandstructure for light in a silicon inverse woodpile photonic crystal with $\frac{r}{a}=0.190$ and $\epsilon_{\rm{si}}=12.1$ for the irreducible Brillouin zone of a simple orthorhombic lattice, calculated with the method of Ref.~\cite{mpb}. This bandstructure appears to be representative for the crystal of Fig \ref{fig1}$(a)$. Between reduced frequency $\frac{a}{\lambda} = 0.395$ ($5727$ cm$^{-1}$) and $\frac{a}{\lambda} = 0.460$ ($6668$ cm$^{-1}$) a broad photonic band gap appears with $15.1\%$ relative width (blue bar). The bandstructure in both the $\Gamma X$- and $\Gamma Z$-direction is identical and shows two stopgaps, marked by the grey rectangles in the $\Gamma Z$-direction. The lowest frequency stopgap between $\frac{a}{\lambda} = 0.310$ ($4505$ cm$^{-1}$) and $\frac{a}{\lambda} = 0.318$ ($4611$ cm$^{-1}$) is narrow and it is closed when going in the $ZU$-direction. The broad second stopgap between reduced frequency $\frac{a}{\lambda} = 0.395$ ($5727$ cm$^{-1}$) and $\frac{a}{\lambda} = 0.488$ ($7062$ cm$^{-1}$) is part of the photonic band gap. The low frequency edge of the photonic band gap is bounded along the $\Gamma X$- and $\Gamma Z$-direction, which we observe in our experiment. The high frequency edge of the band gap occurs at the $S$ and $T$ points, which are not accessible in our experiment. Deviations in the crystal geometry affect the dispersion relation and therefore the band gap \cite{Hillebrand2004, Woldering2009}. The band gap of this kind of structure is robust to most types of deviations, whose tolerances are well within reach of our fabrication methods \cite{Woldering2009}.

\section{Experimental setup}

A supercontinuum white light source (Fianium SC-450-2) with a frequency range of $4000$ to $22000$ cm$^{-1}$ is used to illuminate the photonic crystals. Light is polarized and focused on the crystal using a gold-coated reflecting objective to avoid dispersion (Ealing X74). The numerical aperture NA=0.65 of the objective results in a spectrum angle-averaged over approximately $0.44 \pi \pm 10 \%$ sr solid angle in air. The diameter of the focused beam is estimated to be $2w_0 \approx 1$ $\mu$m from experiments on micropillars \cite{Ctistis2010}. Reflectivity is measured for a broad range of wavevectors that are centred on the $\Gamma X$- and $\Gamma Z$-directions, which is expected to result in similar stopbands because of the symmetry of the crystal. Reflected light is collected by the same objective, and the polarization is analysed. The spectrum is resolved using Fourier transform infrared spectroscopy (BioRad FTS-6000) in combination with an external InAs photodiode. The reflectivity of crystals is collected between $4000$ and $10000$ cm$^{-1}$ with a resolution of $32$ cm$^{-1}$. For calibration, spectra are normalized to the reflectance spectra of a gold mirror, which were collected before and after measuring on a crystal. The individual gold spectra show only minor differences due to the excellent stability of the setup, except between $9200-9600$ cm$^{-1}$, close to the master frequency of the white light source, which is thus excluded. The reflectivity is measured for two orthogonal polarizations, where the orientation of the set of pores perpendicular to the measurement direction is used as reference; light is $\perp$-polarized when the electric field is perpendicular to the direction of this set, light is $\parallel$-polarized when it is parallel with the direction of this set. For example, when the reflectivity is measured centred on the $\Gamma X$-direction (see Fig. \ref{fig1}$(a)$), light is $\parallel$-polarized when the electric field is parallel with the $\Gamma Z$-direction. The samples are placed on a three-dimensional x-y-z-translation stage (precision $\pm 50$ nm) to study the position-dependence of the reflectivity.

\section{Results}

Position-dependent reflectivity experiments have been performed by scanning the photonic crystal through the focal spot. These scans were performed on two orthogonal surfaces of the crystal allowing for measurements along the $\Gamma X$- and $\Gamma Z$-direction, see Fig. \ref{fig1}$(b)$. These scans provide insight in the reproducibility of the reflectivity, the optical size of the crystal, and boundary effects. We will first describe one specific scan for the reflectivity obtained centred on the $\Gamma X$-direction for $\perp$-polarized light. Next we will compare reflectivity spectra obtained from different scans to demonstrate position-independent overlapping stopbands for orthogonal polarizations and crystal directions that are the signature for the photonic band gap.

\begin{figure*}[tbp]
  \includegraphics[width=16 cm]{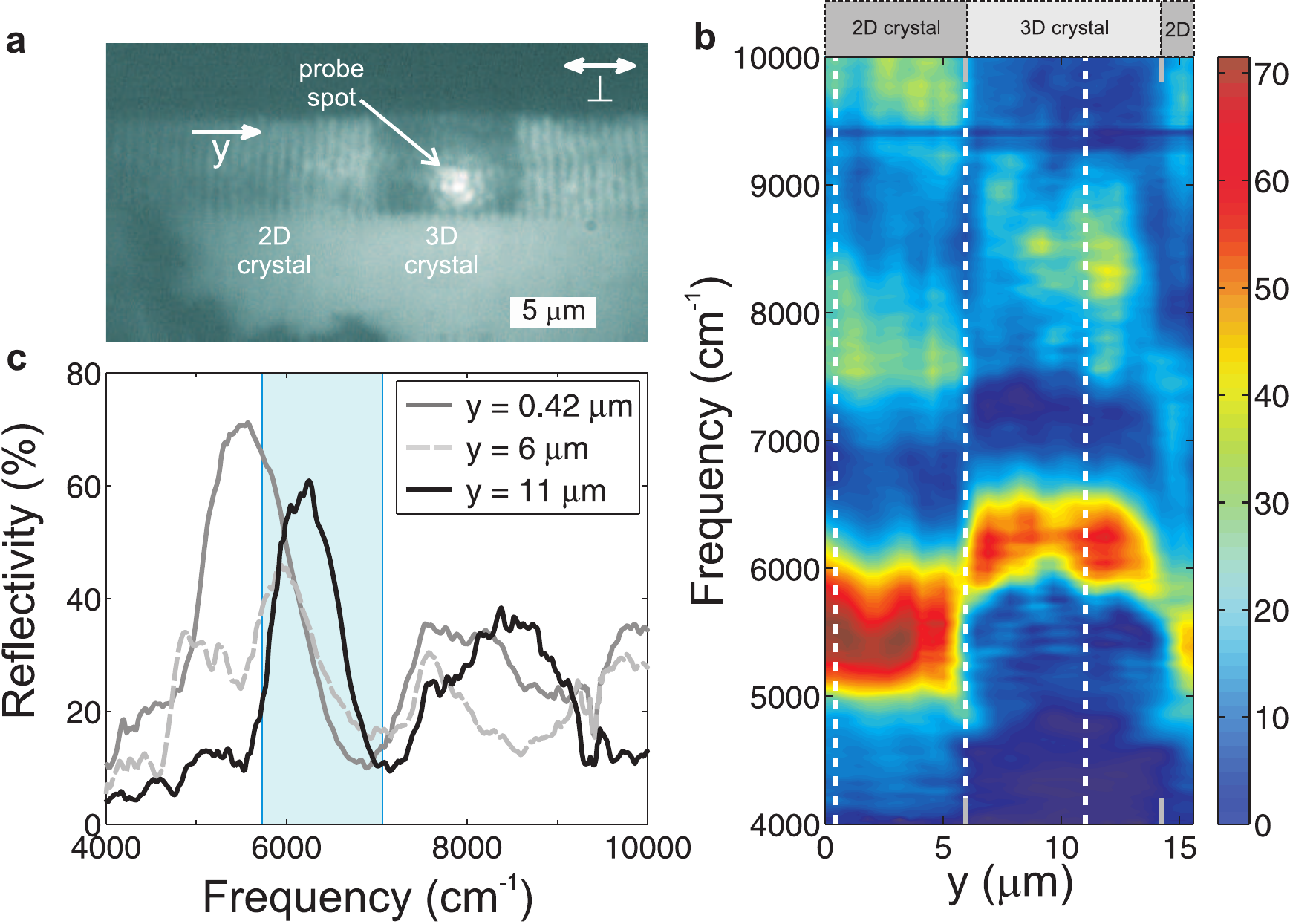}
\caption{\textit{(color)} Position- and frequency-resolved reflectivity for an inverse woodpile photonic crystal, measured parallel with the $\Gamma X$-direction for $\perp$-polarized light. $(a)$ Microscope picture of the crystal placed in the setup. The bright spot near the center of the crystal is reflected focussed light of the supercontinuum white light source. The spot was moved over the crystal surface along the $y$-direction. $(b)$ Reflectivity collected as a function of the $y$-position of the spot on the photonic crystal surface. The scan started on the surrounding two-dimensional crystal ($y=0$ $\mu$m), continued over the inverse woodpile photonic crystal ($6 < y<14$ $\mu$m) and ended on the two-dimensional crystal. The spot was moved in steps of $0.46 \pm 0.05$ $\mu$m. The grey lines on the horizontal axis indicate the boundaries of the crystal. The white dashed lines indicate positions for representative spectra displayed in $(c)$. $(c)$ Reflectivity spectra measured at the $y$-coordinates given by the white dashed lines in $(b)$. The calculated stop gap in the $\Gamma X$-direction for $\frac{r}{a}=0.19$ is marked by the blue bar.}
\label{fig3}
\end{figure*}

\subsection{Position-dependent reflectivity along the $\Gamma X$-direction for $\perp$-polarized light.}

Fig. \ref{fig3} shows the position-dependent reflectivity along the $\Gamma X$-direction for $\perp$-polarized light, where the focal spot was moved in the $y$-direction. Fig. \ref{fig3}$(a)$ shows an optical microscope image of the inverse woodpile photonic crystal (dark square in the middle) taken during the reflectivity scan, viewed along the same perspective as Fig. \ref{fig1}$(c)$. The bright spot centred on the inverse woodpile photonic crystal is the focal spot of the supercontinuum white light source. The scan was started with the focal spot on the surrounding two-dimensional crystal  to the left of the inverse woodpile photonic crystal, continued over the inverse woodpile photonic crystal and finished on the two-dimensional crystal to right of the inverse woodpile photonic crystal. Fig. \ref{fig3}$(b)$ shows a contour-plot of the measured reflectivity as a function of position,  Fig. \ref{fig3}$(c)$ shows three cross-sections of the contour-plot of Fig. \ref{fig3}$(b)$.

In Fig. \ref{fig3}$(b)$ a broad reflectivity peak is observed for the inverse woodpile photonic crystal ($6<y< 14$ $\mu$m) between $5600$ and $6900$ cm$^{-1}$ with a high reflectivity of $60\%$. The spectrum taken at $y=11$ $\mu$m is shown in Fig. \ref{fig3}$(c)$. This distinct peak is observed in the same frequency range over the entire three-dimensional crystal surface. The stopband corresponds to the stopgap in the $\Gamma X$-direction, which is marked by the blue bar in Fig. \ref{fig3}$(c)$, and which is part of the photonic band gap.

The maximum reflectivity is about $60\%$ for our inverse woodpile photonic crystals, which is likely limited by surface roughness that scatters light in directions that are not collected by the objective. This assertion is supported by the observation of lower reflectivity on crystal surfaces that were less well polished. The finite thickness of the crystals could also lead to some reduction of the reflectivity. From the full width at half maximum of the peak, we derive a Bragg length equal to approximately $7d$, with $d$ the spacing between lattice planes \cite{ThesisKoenderink}. Therefore, we conclude that our crystals have a thickness of approximately 1.5 times the Bragg length. This corresponds to a maximum reflectivity of $1-e^{-1.5}=78\%$. Since the observed maximum reflectivity is consistent with surface roughness and finite size effects, it is reasonable that the crystal has a high reflectivity through all of the $0.44 \pi \pm 10 \%$ sr probed solid angle. As is seen in the bandstructures in Fig. \ref{fig2}, the stopgap hardly shifts in frequency for different directions in the Brillouin zone, hence it is even more likely that the obtained angle-averaged reflectivity is representative for the full probed range.

From the scan in Fig. \ref{fig3}$(b)$, we observe a clear distinction between the reflectivity of the three-dimensional inverse woodpile photonic crystal and the surrounding two-dimensional crystal. A reflectivity maximum of $70\%$ is observed on the two-dimensional crystal at $y < 6$ $\mu$m and at $y > 14$ $\mu$m. The spectrum taken at $y=0.42$ $\mu$m is shown in Fig. \ref{fig3}$(c)$. This peak belongs to the $\Gamma K$ stopgap between $4700$ and $6300$ cm$^{-1}$ for TE polarized light \cite{Nair2010}. At the interfaces between the two-dimensional and inverse woodpile photonic crystal at $y = 6$ and at $y=14$ $\mu$m, a peculiar double peak appears as shown in Fig. \ref{fig3}$(c)$, which is only present at the two-dimensional crystal to three-dimensional crystal interface. The spectrum at $y=6$ $\mu$m is shown in Fig. \ref{fig3}$(c)$. This double peak consist of one component between approximately $4700$ and $5500$ cm$^{-1}$, and a second component between approximately $5500$ and $6500$ cm$^{-1}$. This double peak differs from a linear combination of two-dimensional crystal and three-dimensional crystal peaks (see Fig. \ref{fig3}$(c)$) that would be expected naively, since both of its components appear at lower frequencies than the two-dimensional crystal and three-dimensional crystal peaks. While we have currently no interpretation for this intricate double peak, the observation of interface peaks calls for advanced theoretical interpretations outside the scope of the present paper. In the long-wavelength limit below 5000 cm$^{-1}$, both the two-dimensional and inverse woodpile photonic crystals are transparent. As a result, their reflectivities are low. Since the silicon filling fraction is smaller for inverse woodpile photonic crystals than for two-dimensional crystals, they have a lower effective refractive index \cite{Datta1993}, resulting in a lower long-wavelength reflectivity for inverse woodpile photonic crystals.

By combining this $y$-scan with six other scans for orthogonal directions, crystal surfaces, and polarizations, we can measure the optical size of the crystal, in other words, the range over which the crystal has the same reflectivity. This is in the $x$-direction approximately $3$ $\mu$m, in the $y$-direction $7$ $\mu$m and in the $z$-direction $4$ $\mu$m. The optical size is smaller in the $x$-direction than in the $y$-direction, which is in agreement with detailed structural information derived in Ref.~\cite{vandenBroek2011} for inverse woodpile photonic crystals.

\begin{figure}[tbp]
  \includegraphics[width=8.6 cm]{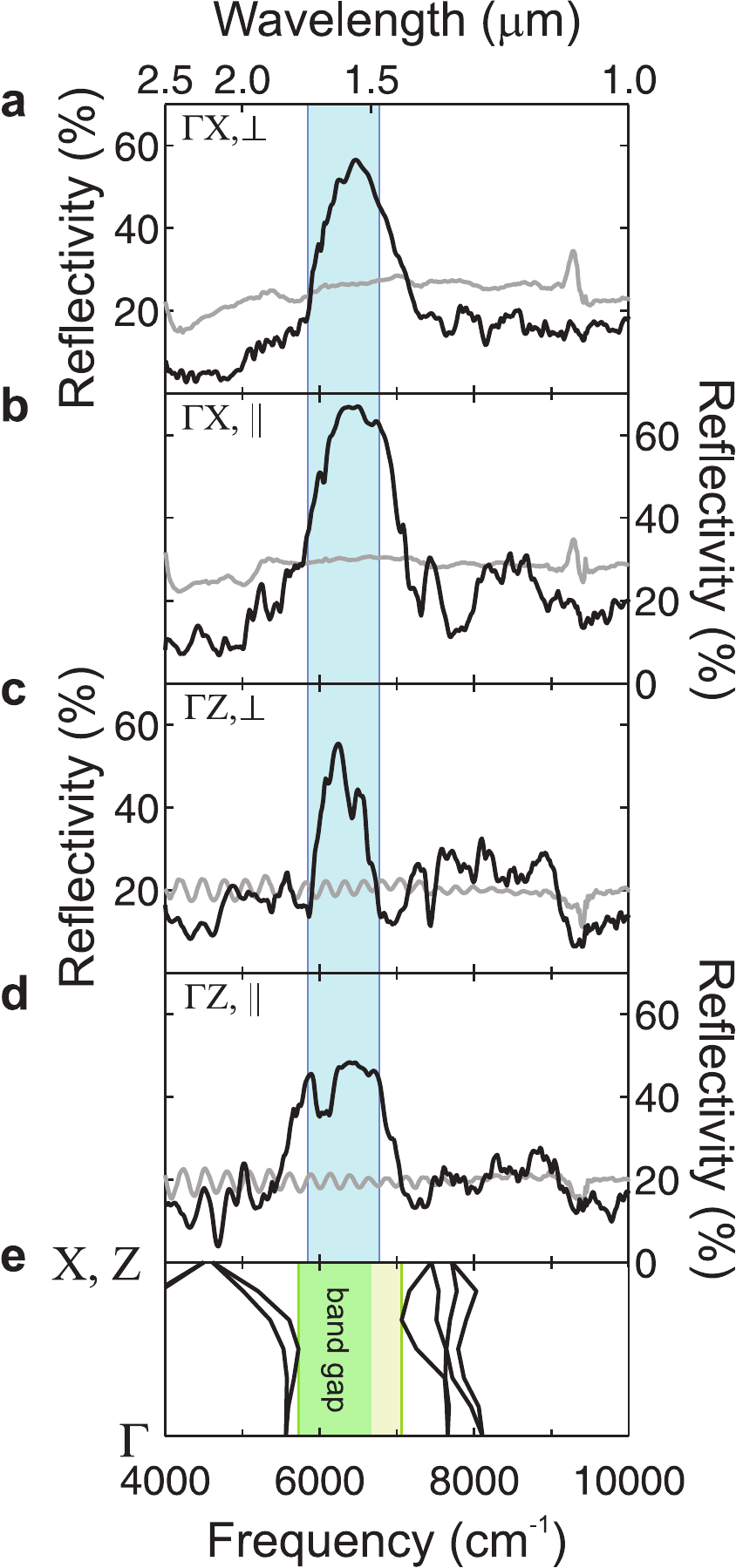}
\caption{\textit{(color online)} $(a-d)$, Measured reflectivity (black lines) for an inverse woodpile photonic crystal along two orthogonal directions (centred on $\Gamma X$ and $\Gamma Z$) and orthogonal polarizations (perpendicular $\perp$ and parallel $\parallel$ to the pores perpendicular to the measurement axes). The overlapping frequencies of the stopbands (blue bar) reveal the presence of a photonic band gap. Reference reflectivity (grey lines) are measured on non-photonic parts of the sample within $5$ $\mu$m from the photonic crystal. $(e)$, Calculated bandstructure for both $\Gamma X$ and $\Gamma Z$ reveal a stopgap (light green bar), which is part of the photonic band gap (dark green bar). }
\label{fig4}
\end{figure}

\subsection{Overlapping stopbands for orthogonal polarizations and crystal surfaces.}

We have studied the position-dependent reflectivity spectra for two orthogonal crystal surfaces and two orthogonal polarizations, by taking scans similar to the one presented in Fig. \ref{fig3}. From these scans we have selected 4 $(=2\times2)$ spectra that are reproducible on several locations on the crystal, see Fig. \ref{fig4}$(a-d)$, black curves. Fig. \ref{fig4}$(a)$ shows the reflectivity centred in the $\Gamma X$-direction for $\perp$-polarized light, and Fig. \ref{fig4}$(b)$ for $\parallel$-polarized light, Fig. \ref{fig4}$(c)$ shows the reflectivity centred in the $\Gamma Z$-direction for $\perp$-polarized light, and Fig. \ref{fig4}$(d)$ for $\parallel$-polarized light. A broad reflectivity peak appears for both directions and polarizations with maxima up to $67\%$. The stopbands overlap between $5900$ and $6900$ cm$^{-1}$ (blue bar), corresponding to a relative bandwidth of $16\%$. This position-independent overlapping stopband for orthogonal polarizations and crystal directions is a signature for the photonic band gap with a relative bandwidth up to $16\%$.

The reflectivity obtained from the $\Gamma X$-direction is shown in Fig. \ref{fig4}$(a-b)$. Below $5000$ cm$^{-1}$, the reflectivity behaves similarly for $\perp$- and $\parallel$-polarized waves. Based on the filling fraction of this crystal, one would expect a reflectivity of approximately $10\%$, which is in very good agreement with our observations. A major reflectivity peak is observed for both polarizations; for $\perp$-polarized light (Fig. \ref{fig4}$(a)$) the peak appears between $5800$ and $7300$ cm$^{-1}$, for $\parallel$-polarized waves the peak appears between $5300$ and $7100$ cm$^{-1}$. At higher frequencies the reflectivity varies around $20\%$. Reference spectra (grey) were collected on the silicon surface within $5$ $\mu$m distance from the inverse woodpile photonic crystal. The reflectivity varies between about $20$ and $30\%$, which agrees closely with $31\%$ reflectivity based on Fresnel reflection and a refractive index of $n=3.5$ for a flat silicon surface. The reference spectra confirm that the observed stopbands in the inverse woodpile photonic crystal are indeed caused by the geometry of the nanostructure.

The reflectivity obtained from the $\Gamma Z$-direction is shown in Fig. \ref{fig4}$(c-d)$. Below $5000$ cm$^{-1}$, the reflectivity behaves similarly for $\perp$- and $\parallel$-polarized waves. For the $\Gamma Z$-direction, fringes appear and the average reflectivity is higher than for the $\Gamma X$-direction. These fringes are Fabry-Perot interferences caused by internal reflections in the inverse woodpile photonic crystal between the air-crystal boundary and the crystal-silicon boundary. The fringes are also present on the non-photonic reference spectra (grey) that are collected parallel to the first set of pores that form the surrounding two-dimensional crystal. Above $5000$ cm$^{-1}$ a  major reflectivity peak is observed for both polarizations near the expected band gap. For $\perp$-polarized light (Fig. \ref{fig4}$(c)$) the peak appears between $5900$ and $6900$ cm$^{-1}$, for $\parallel$-polarized waves the peak appears between $5300$ and $7200$ cm$^{-1}$. The amplitude of the reflectivity peaks in the $\Gamma Z$-direction are comparable with the reflectivity peaks of the $\Gamma X$-direction. For the $\Gamma Z$-direction, several troughs appear in the spectra, which are likely caused by surface contamination.

If one overlays the spectra for $\perp$-polarized light, compare Fig. \ref{fig4}$(a,c)$, it appears that the peaks overlap at the low-frequency side. In addition, for both the $\Gamma X$- and $\Gamma Z$-directions, the stopband is broader for $\parallel$-polarized waves than for $\perp$-polarized waves. This is an experimental indication that the pore radii of both sets of pores are identical, as expected from the symmetry of the structure.

\begin{figure}[tbp]
  \includegraphics[width=8.6 cm]{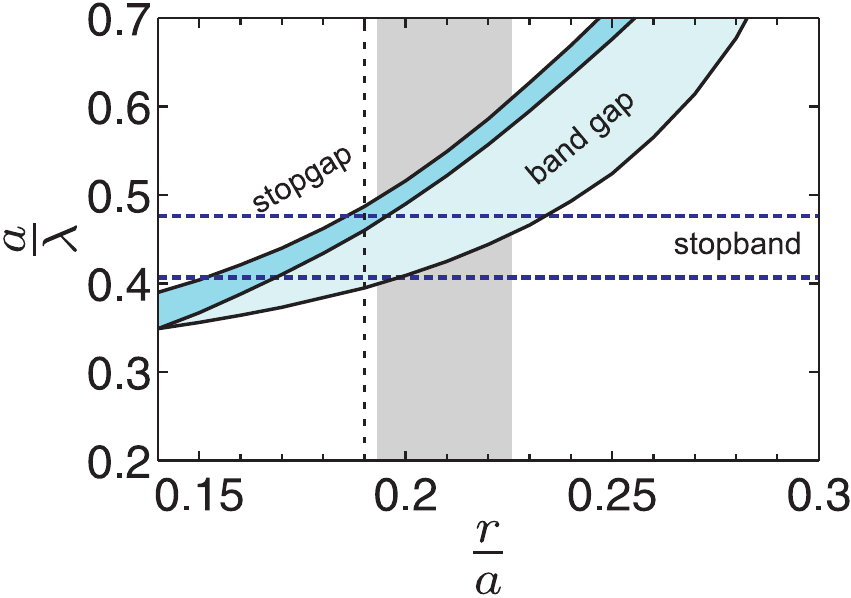}
\caption{\textit{(color online)} Calculated stopgap for the $\Gamma X$- and $\Gamma Z$-directions as a function of relative pore radius $\frac{r}{a}$. The band gap (light blue) is part of the broad stopgap (dark blue). The grey bar is the value of $\frac{r}{a}$ based on parameters determined from SEM images. The bandstructures in Fig. \ref{fig2} and \ref{fig4}$(e)$ were calculated at $\frac{r}{a}=0.19$ (vertical dashed line) that agree best with the experiments. The horizontal dashed lines mark the edges of the common stopband determined from Fig. \ref{fig4}$(a-d)$.}
\label{fig6}
\end{figure}

\section{Discussion}

In order to interpret our observations, notably of the overlapping stopbands for two orthogonal crystal directions and two orthogonal polarizations, we have performed photonic bandstructure calculations. In the course of the calculations, we have come to realize that pore radii determined from electron microscopy are an overestimate of the true pore radii. This is reasonable, since parameters determined from electron microscopy have un uncertainty, typically near 2 to 3 $\%$ \cite{Megens1997}. Therefore, we have performed bandstructure calculations as a function of $\frac{r}{a}$. Fig. \ref{fig6} shows the calculated stopgap in the $\Gamma X$- and $\Gamma Z$-direction and the band gap versus $\frac{r}{a}$. The broad stopgap corresponds to the intense stopbands in our reflectivity spectra. The horizontal dashed lines mark the edges of the overlapping stopband observed in Fig. \ref{fig4}. The grey bar is the value of $\frac{r}{a}$ based on parameters determined from SEM images. For a relative pore radius between $\frac{r}{a}=0.19$ and $\frac{r}{a}=0.20$, the observed stopband is in agreement with the calculated band gap and stopgap, which is close to the value of $\frac{r}{a}$ based on SEM images. This indicates that the crystal has an $\frac{r}{a}$ between $0.19$ and $0.20$. The corresponding bandstructure for $\frac{r}{a}=0.190$ is shown in Fig. \ref{fig2} and the bandstructure in the $\Gamma X$- and $\Gamma Z$-directions are shown in Fig. \ref{fig4}$(e)$. A stopgap is predicted between $5727$ and $7061$ cm$^{-1}$, which is part of the calculated band gap between $5727$ and $6668$ cm$^{-1}$. The calculated band gap supports the observed signature of the photonic band gap between $5900$ and $6900$ cm$^{-1}$.

We have performed reflectivity measurements with an objective with NA$=0.65$, resulting in a spectrum angle-averaged over $0.44 \pi \pm 10 \%$ sr solid angle in air. We have measured the reflectivity from two orthogonal crystal surfaces and we observe intense peaks. Therefore, as argued above from the high reflectivity and the finite size of the photonic crystals, it is most likely that the observed stopbands extend over $0.88 \pi \pm 10 \%$ sr solid angle. By invoking the symmetry of the crystal, the stopbands centred along the opposite $-\Gamma X$- and $-\Gamma Z$-directions are identical to the $\Gamma X$- and $\Gamma Z$ directions. Therefore, our observations would correspond to stopbands up to $1.76 \pi \pm 10 \%$ sr external solid angle of the inverse woodpile photonic crystal, in other words nearly half of all directions. In previous experiments on inverse opals without bandgap \cite{Thijssen1999}, nearly half of all directions were found to be excluded for propagation for a vanishingly narrow frequency band. Therefore, the present result is to the best of our knowledge the largest solid angle for which a broad photonic stopband has been reported.

The observed common stopband between $5900$ and $6900$ cm$^{-1}$ is well explained with bandstructure calculations. A limitation to our interpretation is that we invoke a theory for infinite crystals. Theoretical calculations of reflectivity for finite crystals could result into new insights in the obtained spectra, for instance how light propagates inside photonic band gap crystals, which is an extremely complex problem as described in Ref. \cite{Notomi2000}. A well known method for finite crystals are finite-difference time domain simulations \cite{Yee1966}. Unfortunately, however, full three-dimensional simulations with reasonable parameters for our experiments will require very extensive numerical calculations \cite{Kole2003} that are outside the scope of our present experimental study. Scattering matrix methods \cite{Li1996, Li1997} provide a good description of three-dimensional crystals with finite thickness. However, the method assumes that a crystal is infinitely extended in both transverse dimensions which is not the case for our crystals. Hence we propose that a theory for three-dimensional finite crystals is called for. It will be even more difficult to implement the experimental boundary conditions into such calculations, such as the shape of the pores and the NA of the microscope objective. At any rate, the agreement between our calculations and our experimental data is gratifying.

For good measure, the observed signature of the photonic band gap is not a complete experimental observation of such a gap. At this time, we can not experimentally exclude that there are crystal directions where the stopbands are either closed or shifted to different frequencies. While we consider these possibilities unlikely based on the theoretical study of our crystals and in view of the measured crystal parameters, a fully experimental observation of a photonic band gap must entail an experimental method where the local density of optical states is probed. An example would be the doping of the crystals with quantum dots \cite{Leistikow2009} and measuring the emission rates \cite{Koenderink2003, Lodahl2004, Ogawa2004}.

\section{Conclusions}
In conclusion, we have studied the polarization-dependent reflectivity centred on the $\Gamma X$- and $\Gamma Z$-direction of three-dimensional silicon inverse woodpile photonic crystals. We have observed a position-independent overlapping stopband for different polarizations and directions, which are supported with calculated bandstructures. This is the signature of a photonic band gap with a bandwidth up to $16\%$.
\section*{Acknowledgments}
We kindly thank Willem Tjerkstra and Johanna van den Broek for their expert efforts in sample fabrication and preparation, and Alex Hartsuiker for helpful discussions. This work was supported by the Stichting Fundamenteel Onderzoek der Materie (FOM) that is financially supported by the Nederlandse Organisatie voor Wetenschappelijk Onderzoek (NWO). WLV also thanks NWO-VICI, STW-NanoNed, and Smartmix-Memphis.


\begin{thebibliography}{10}
\bibitem{CallibrationErrors}
These parameters do not include possible calibration errors of the scanning electron microscope.
\end{thebibliography}
\end{document}